\documentclass[letterpaper,twocolumn,10pt]{article}
\usepackage{usenix2025_SOUPS}

\usepackage{tikz}
\usepackage{amsmath}
\usepackage{xurl}
\usepackage{booktabs}
\usepackage{longtable}

\renewenvironment{quote}
  {\list{}{\rightmargin=0em \leftmargin=1em}%
   \item\relax}
  {\endlist}

\begin{document}

\date{}

\title{\Large \bf Minoritised Ethnic People's Security and Privacy Concerns and Responses towards Essential Online Services}

\def\plainauthor{Aunam Quyoum, Mark Wong, Sebati Ghosh and Siamak F.\ Shahandashti}

\author{
{\rm Aunam Quyoum} \\ University of Glasgow
\and
{\rm Mark Wong} \\ University of Glasgow
\and
{\rm Sebati Ghosh} \\ University of York
\and
{\rm Siamak F.\ Shahandashti} \\ University of York
} 

\maketitle
\thecopyright

\begin{abstract}
Minoritised ethnic people are marginalised in society, and therefore at a higher risk of adverse online harms, including those arising from the loss of security and privacy of personal data. Despite this, there has been very little research focused on minoritised ethnic people's security and privacy concerns, attitudes, and behaviours. In this work, we provide the results of one of the first studies in this regard. We explore minoritised ethnic people's experiences of using essential online services across three sectors: health, social housing, and energy, their security and privacy-related concerns, and responses towards these services. We conducted a thematic analysis of 44 semi-structured interviews with people of various reported minoritised ethnicities in the UK. Privacy concerns and lack of control over personal data emerged as a major theme, with many interviewees considering privacy as their most significant concern when using online services. Several creative tactics to exercise some agency were reported, including selective and inconsistent disclosure of personal data. A core concern about how data may be used was driven by a fear of repercussions, including penalisation and discrimination, influenced by prior experiences of institutional and online racism. The increased concern and potential for harm resulted in minoritised ethnic people grappling with a higher-stakes dilemma of whether to disclose personal information online or not. Furthermore, trust in institutions, or lack thereof, was found to be embedded throughout as a basis for adapting behaviour. We draw on our results to provide lessons learned for the design of more inclusive, marginalisation-aware, and privacy-preserving online services.
\end{abstract}

\section{Introduction}

Online systems are promising innovations in digital services and often praised for their ability to offer the public greater control and added convenience in managing their day-to-day lives. Despite the benefits, user-centred research in various fields of enquiry has shown the public remain cautious and fearful about the effects of increased digitalisation on privacy and security~\cite{dencik2018data}. Data privacy concerns in this context include apprehension about the excessive collection, aggregation, sharing, and use of personal information. Such concerns are interwoven with security concerns which range from worrying about loss of access to one’s data, services and devices, to the fear of confidential data about one’s affairs being compromised. These concerns are often exacerbated by usability and accessibility issues within online services. This results in some people becoming reluctant to engage with online services, and in some cases choosing to withdraw from engaging with services entirely as they become ‘digital by default’~\cite{lyu2024understanding}.  

Minoritised ethnic people are one of the groups that emerging evidence suggests are refusing and withdrawing from online services due to prior experiences and concerns related to racism, lack of trust towards institutions and digitalisation, and privacy and security concerns~\cite{quyoum2024valuing}. However, there is still a lack of understanding in the literature on how experiences of “race”, racism, and ethnicity affect people’s privacy and security-related attitudes and behaviours. It is therefore important to understand why particular users are concerned about the risks of harm, or refuse adoption and use of technology altogether, in order to protect their privacy and security. This will help service providers, designers, developers, and policymakers gain a better understanding of the underlying factors affecting users’ confidence and trust in essential online services and ensure their designs are user-centred.  

We note that this paper uses the term ‘minoritised ethnic’ people, as it is more commonly used in the UK context by communities, policymakers, and third sector as a generalised term to refer to people who are often adversely racialised by social processes of power. Racialised people are historically and structurally ‘minoritised’ and are oppressed by systems of power that perpetuate inequity and subjugation, based on their skin colour, cultural or religious practice~\cite{murji2017racism}. This term recognises that individuals have been minoritised, as opposed to being just distinct statistical minorities, as many minoritised ethnic people in the UK form part of wider global majorities. We acknowledge the limitations of a single term however, and seek to avoid homogenising the differences within and between peoples' ethnicities, heritages, and identities. However, we recognise there are commonalties and shared experiences of being racialised, marginalised, and “other-ed” by ideologies of “race” and racism in the UK context. 

In privacy and security research, there has been acknowledgement of the need to consider the experiences of marginalised users when designing digital services~\cite{walker2019moving}. In the same vein, Wang argues for ‘inclusive privacy’, which centres the design process around users with various characteristics, abilities, needs, and values, who are often under-served~\cite{wang2018inclusive}. There have been several works studying users marginalised by their disability, gender, and sexual orientation. Privacy and security concerns and behaviours of people with visual impairment and the lack of technology accessibility to address such concerns is one of the better studied areas in this field (see, e.g., \cite{ahmed2015privacy}, \cite{dosono2015m}, and \cite{hayes2019cooperative}). Another group that has received some research attention in this area is people with cognitive impairment (see, e.g., \cite{cornejo2016vulnerability} and \cite{mentis2020illusion}). Groups marginalised due to their gender identity and sexual orientation have been the subject of multiple studies as well (see, e.g., \cite{devito2018too}, \cite{maris2020tracking}, \cite{lerner2020privacy}, and \cite{coopamootoo2023equal}). However, minoritised ethnic users are particularly understudied in this regard, but necessary to include for the purpose of inclusive design and research.   

Furthermore, there is currently a lack of understanding of how the social context of data and the experience of racism interacts with attitudes towards privacy and security of digital technology. This is an important gap to address because the negative impact of privacy and security issues are arguably more severe for minoritised ethnic people who are already marginalised and racialised in society. Thus, any unwanted leakage of personal data such as ethnicity, or proxies for ethnicity, e.g., name, may lead to more adverse harms and further social and economic exclusion that exists because of systemic racism and marginalisation. Srinivasan et al.\ highlight that marginalised groups face disproportionate barriers because of varied levels of digital literacy, access to devices, linguistic or cultural differences, 
and skills and confidence of using technology to achieve specific tasks~\cite{srinivasan2018privacy}. Islam et al.\ echo many of these issues in the context of digital healthcare in the UK and find digital precarity, including inadequate access to devices, internet connectivity and digital literacy skills as major barriers for minoritised ethnic people~\cite{islam2024digitalised}. 

There is also an empirical gap, as minoritised ethnic people are an understudied cohort in privacy and security research. More broadly, across disciplines where research with such groups has been undertaken, existing studies often incur challenges in terms of recruitment and accessibility of language translation. But studies that focus on the impact and experiences of systemic racism, and how it affects people’s interactions with technology, are rare. Moreover, while existing studies typically provide some general understanding through survey approaches, they unfortunately offer less nuanced insight into people’s attitudes, experiences, and behaviours.  

In this paper, we offer an in-depth analysis of data collected through individual interviews with 44 minoritised ethnic people in the UK. This addresses important and empirical gaps as few existing studies have focused on the lived experiences of minoritised ethnic communities in the UK. The research questions guiding this focused analysis were to examine the security and privacy related \emph{concerns}, the behavioural \emph{responses} to these concerns, and the \emph{root causes} of such concerns among minoritised ethnic people interacting with essential online services. We were especially interested in how concerns and responses may be relevant to being a racialised minority.   

Our approach to this study draws on experience in social science research, particularly on racism, lived experience of minoritised ethnic people, and use of qualitative and participatory research. This is combined with experience in security and privacy research in computer science, especially human factors in security and privacy, to develop this paper. This collaborative, cross-disciplinary approach allowed the authors to draw from a richer repertoire of literature and aim to provide a more holistic analysis which centres the lived experience of people who are ethnically minoritised in society. 

Our study focuses on minoritised ethnic people’s experiences of online services in three specific sectors in the UK, and particularly in England and Scotland: health, social housing, and energy. These represent three essential services that vastly differ in their delivery model. The healthcare system in the UK, known as the National Health Service (NHS), constitutes a public service that is delivered at the national level (particularly relevant to our study are NHS England and NHS Scotland). Healthcare services have been severely underfunded in the past few decades and inequalities in the UK have been deepened in this context, leading to increasing levels of poverty and destitution, of which some minoritised ethnic groups, disabled, and social housing renters are facing higher rates~\cite{rowntree2025poverty}. Social housing, while still a public service, is managed at the local government level and hence service practices vary greatly between different areas. The energy ecosystem is quite complex, but for consumers, although there are measures by the government to control prices, the provision of energy services, including gas and electricity, is privatised in the UK. All three sectors are essential to people's livelihoods, making it all the more pertinent to investigate as these services are increasingly digitalising aspects of their delivery. 

Yet, people’s online experiences, sense of trust, and fear of security and privacy-related harms cannot be necessarily compartmentalised based on \emph{which} service is being provided, as it is still situated within a wider social context where power, racial profiling, bias, and discrimination is persistent. 
For minoritised ethnic people, the potential for digital discrimination is concerning and was expressed by participants in this research. 
We also discuss wider themes related to trust to argue that what happens offline and prior experiences of racism and discrimination can influence perceptions of trustworthiness and affect online behaviours.

\section{Literature Review}

Given the disproportionate harm marginalised people experience when their privacy is violated, a growing need has been felt to represent their voice and perspective in security and privacy research. However, the literature remains limited in understanding the experiences of such groups. 
We examine some of the related work that considers security and privacy concerns and practices of marginalised groups, specifically focusing on literature examining ethnicity as a direct or indirect marginalisation factor. Recognising the inherent cross-disciplinary nature of the topic, we review literature across various disciplines in computer and social sciences.

\paragraph{Unequal Privacy.}

Several studies provide evidence for an ``online privacy divide'', showing significant disparities in privacy concerns and behaviours between different socioeconomic and demographic groups (see, e.g.,~\cite{dodel2023inequalities}), including between people of different ethnic backgrounds~\cite{madden2017privacy}. The root causes of this divide based on ``race'' have been examined by Reichel~\cite{Reichel2017race} and Fullenwieder and Molnar~\cite{fullenwieder2018settler}. 


Reichel’s examination of privacy rights, “race”, and class critically analyses how privacy rights are distributed unevenly across “race” and class in capitalist societies~\cite{Reichel2017race}. By examining examples such as disparities between encryption options across devices, the author argues privacy rights reflect and further reinforce existing power dynamics in society, which serve the interests of dominant, privileged social groups, while others, especially racially and economically marginalised groups, are subjected to systematic surveillance. Hence, showing that “race” plays a crucial role in how privacy is distributed. 


Fullenwieder and Molnar critique the notion of privacy rights for collectives of people who experience systemic injustice~\cite{fullenwieder2018settler}. 
The authors critique liberal privacy frameworks, arguing how privacy can be used to facilitate  systemic violence under the guise of protecting survivors’ confidentiality. While the notion of individual privacy rights is empowering, these frameworks are argued to be ill-suited for addressing collective indigenous experiences and histories. 


The limitations of individualistic privacy notions have been further highlighted by others. Masur et al.\ argue that individualised conceptualisations often view people uniformly, ignoring their social, political, and economic power differences, whereas structurally marginalised groups, e.g., people of colour, immigrants, and religious minorities, experience privacy differently from privileged individuals~\cite{masur2025comparative}. This unequal experience of privacy is also discussed by Gangadharan who argues that digital inclusion without privacy safeguards can reinforce historical patterns of inequality, particularly for marginalised communities~\cite{gangadharan2017downside}. Gangadharan argues that historically marginalised groups, particularly people of colour, immigrants, and low-income individuals, are disproportionately subjected to various types of digital oppression through privacy invasions, like data profiling, digital tracking, and surveillance, both by governments and corporations. Srinivasan et al.\ echo similar findings comparing the expectation and experience of privacy between marginalised and privileged groups, and argue privacy scholarship must expand beyond its traditional focus on privileged, Western subjects and incorporate the unique concerns and experiences of marginalised communities, including people of colour~\cite{srinivasan2018privacy}. 

\paragraph{Interplay between Ethnicity and Privacy.}

Other authors have examined specific communities and, although this was not their primary focus, they observed an interplay between racial or ethnic identity and privacy and security-related concerns and behaviours. For instance, Van Staden \& Bidwell discuss how structural inequalities, historical legacies, and social power dynamics affect privacy governance in Namibia~\cite{vanstaden2024localised}; Wang \& Metzger examine the nuanced relationship between “race”/ethnicity and privacy concerns and management strategies in the context of social media in the US~\cite{wang2024online}; Kam et al.\ discuss Latinx undocumented immigrants’ disclosure practices in the US~\cite{kam2020latinx}; and Simko et al.\ highlight how refugees’ racial, ethnic, and cultural backgrounds influence their privacy concerns and practices, and how their experiences of authoritarian surveillance, digital illiteracy, and systemic discrimination shapes their relationship to online privacy~\cite{Simko2018refugees}. 

The effect of cultural and social norms on security and privacy concerns and behaviour has been explored in existing research. Ahmed et al., for example, discuss several social and cultural norms which shape digital privacy concerns in the `Global Majority', focusing on families in Bangladesh as a case study~\cite{ahmed2017digital}. Such norms included those related to patriarchy, power dynamics, and economic resources. Rennie et al.\ focus on how privacy frameworks, designed for individual device use, interact with Aboriginal cultural norms of relatedness and demand sharing, and explore how cultural dynamics affect privacy management, leading to issues such as unauthorised access to banking accounts, digital exclusion, and even community conflict~\cite{Rennie2018privacy}. 
While these studies have looked at contexts linked to “race” or ethnicity, e.g., refugees and undocumented immigrants, they did not, however, explicitly examine the role of “race” in relation to privacy-related concerns or experiences (see, e.g., \cite{guberek2018keeping}). 

\paragraph{The Interplay in Specific Technological Contexts.}

The intersection of “race”/ethnicity and privacy-related concerns and behaviours with respect to specific technologies has received some research attention. Atienza et al.\ consider the role of “race”/ethnicity in shaping attitudes and perceptions towards the use of mobile health (mHealth) technology~\cite{atienza2015consumer}. They found ‘significant diversity in attitudes regarding mHealth privacy/security both within and between traditional demographic groups’~\cite[p.\,1]{atienza2015consumer}. Forte et al.\ explore how the contributors to open collaboration projects sometimes face discrimination due to their ethnic backgrounds and how this, in turn, influences their privacy practices and concerns~\cite{forte2017privacy}. They note that some Wikipedia editors and contributors experienced threats or challenges due to their ethnicity and identity, rather than content quality. They discuss \emph{privilege} in relation to online safety, where some participants acknowledged that their “race”, gender, or nationality made them less vulnerable to privacy risks compared to others. Anthony et al.\ also discuss how “race” and ethnicity influence privacy-related matters through the disproportionate surveillance on marginalised groups, reduced control over personal data, and systemic discrimination through AI and big data, and this in turn contributes to broader social inequalities~\cite{anthony2017toward}. 

\paragraph{Summary.}

Overall, privacy and security research has not adequately explored racial and minoritised ethnic experiences, despite the known issues related to surveillance, bias, and systemic discrimination. This view is also echoed by Sannon and Forte who call for studies that specifically centre “race” to fill this critical gap in the literature~\cite{Sannon2022privacy}. Furthermore, frameworks that are heavily used in the literature to analyse security and privacy-related concerns and behaviours, e.g., the MITRE ATT\&CK taxonomy of cyberattack lifecycle (see \url{attack.mitre.org}) and Solove’s taxonomy of privacy~\cite{solove2006taxonomy}, usually focus on information and how it flows through systems, and not necessarily how the individual interacts with these systems in the wider social context. People in many of these frameworks are considered ‘subjects’ and their agency is often not acknowledged. This paper is positioned, therefore, to follow a user-centred perspective and highlights the daily dilemmas minoritised ethnic people face regarding their privacy and security, and the choices they make in response. 

\section{Methodology}
Our paper draws on qualitative data collected from a wider research project investigating online services in health, social housing, and energy in the UK. The project examined minoritised ethnic people's wider experiences and concerns, with the aim of co-designing a code of practice for equitable digital services~\cite{wong2024minoritised} and developing socio-technical tools and prototypes, including privacy-enhancing and harm-reduction technologies~\cite{prime-technical-toolkit}. The project employed a range of methods including co-design workshops, interviews, surveys, data modelling, user experience research, and software prototyping and design. This was underpinned by an interdisciplinary approach, in partnership with community and third sector organisations, to understand people's needs. The project subsequently leveraged these insights to co-design tools and frameworks, which centre the voice and lived experiences of minoritised ethnic people. The work presented here is based on the interviews carried out in this wider project. 

Our study draws on a qualitative dataset of individual interviews with 101 minoritised ethnic people aged over 18, living in England and Scotland. While the wider research project investigated more general experiences and concerns in relation to the increasing digitalisation of essential services, specifically in the three sectors, during interviews it was apparent that privacy and security concerns were forefront in the minds of a significant number of interviewees. From 101 interviews, a sub-sample of 44 were selected for analysis based on their substantive discussion of privacy and security concerns. 
More specifically, we required that interviewees discussed a security or privacy-related issue and elaborated on a relevant concern or behaviour.
Interview questions and prompts were primarily designed to elicit discussion around general experiences in accessing online services across the three sectors, the challenges incurred, and associated coping mechanisms (see Appendix~\ref{sec:interview-questions} for list of interview questions).   
Our focus here is on interpreting how interviewees' experiences and perceptions of online services interact, or intersect, with their security and privacy-related concerns and behaviours.  

The research and collection of interview data was approved by the Ethics Committee of Heriot-Watt University, the lead university of the project. All interviewees agreed to direct quotes or interview notes, whichever applicable, being used anonymously for publications. 

Interviews took place in four case study sites known for their ethnic and cultural diversity and established links to local community organisations: Glasgow, Manchester, Bradford, and London. We attempted to reach people of minoritised ethnicities that reflect some of the largest populations in the UK according to national census data – these include African, Bangladeshi, Caribbean, Chinese, Indian, Pakistani, and Mixed Heritage. While the sample cannot be considered wholly representative of the diverse minoritised ethnic population in the UK, the sample does encompass a cross-section of characteristics, including age, ethnicity, disability status, extent of online usage, and experiences of social housing.  

Participants were recruited through the support of community organisations who, as collaborative partners to the project, were directly funded to support participant recruitment. All four partner organisations support racialised groups in the community or advocate for the rights of marginalised groups. For example, Council of Ethnic Minority Voluntary Organisations (CEMVO) is a national intermediary that builds the capacity and sustainability of Scotland’s ethnic minority voluntary and community sector. Caribbean \& African Health Network (CAHN) focuses on health research and policy advocacy. A full list of partner organisations comes in Appendix~\ref{sec:community-org-partners}.  

Semi-structured interviews were conducted to facilitate space for open discussion and allow conversations to be directed by issues most important to participants. Participants were given the choice of in-person or online interview, and the option to have interviews conducted in another language alongside a translator. Participants were also provided the option to refuse the audio recording of interviews, which some of them chose and only notes were taken by the interviewer in lieu of recording. Empathy and care were of key importance during interviews and was shown by ensuring flexibility. For example, venues for interviews were mutually decided based on the time and location most accessible and convenient to interviewees, services and helpful information to support individual circumstances were signposted where relevant, and participants were compensated for their time through cash administered through local community organisations. 

Interview data for this study was analysed by all four authors, applying a \emph{thematic analysis}~\cite{braun2021thematic}. Recognising that discussions of security and privacy-related concerns and behaviours could present at any point during the interview (and the semi-structured nature of the interviews allowed for this flexibility), we did not focus on responses to specific interview questions, and considered interviews holistically. We conducted our analysis in two rounds. In the first round, each author was allocated a set of the 101 interviews and undertook independent initial coding to identify security and privacy related themes and discount the interviews that did not include substantive discussions of security and privacy concerns. In the second round, the authors presented and collaboratively agreed on principal themes and sub-themes recurring across interviews in multiple sittings. This iterative approach of analysis involved comparing themes across the team's analysis to ensure consistency, identify trends, and cross-check understanding of key theoretical concepts.

\paragraph{Positionality Statement.}
Thematic analysis places importance on researcher reflexivity as part of the analytical process~\cite{braun2023toward}. Srinivasan et al.\  recommend discussing positionality as best practice in privacy-related research~\cite{srinivasan2018privacy}, which is relevant to both teams who collected and analysed the data for this paper. Both teams included four researchers who all identify as minoritised ethnic people, which meant they shared some similar experiences of marginalisation, being racialised by institutions and wider cultural references drawn from similar experiences of migration, multilingualism, religion, and from the Global Majority. This helped facilitate a sense of rapport and space for frank discussion during the interviews in particular. However, this is also not to discount the position of power and privilege of researchers employed in higher education institutions, who are not experiencing the same urgency of problems many participants faced in terms of access to social benefits, healthcare, and the cost-of-living crisis.

\section{Analysis and Results}

This section first presents an overview of the sub-sample interviewee demographics and then discusses three overarching themes emerging from data analysis and found to be prominent across a significant number of interviews in the sub-sample. These themes include control over personal data and privacy concerns, dilemma of disclosing personal information, and fear of repercussion. 
For each theme, we elaborate on the relevant security and privacy-related \emph{concerns}, the \emph{behaviours} they lead to, and the \emph{context} (or equivalently \emph{root causes}) that shapes these concerns based on the themes emerging from the interview analysis. Note that the focus of the study was UK health, social housing, and energy services, collectively referred to as ``essential services'' here.
An overarching theme across the data indicated that trust is an inescapable issue and embedded throughout the themes presented. Where applicable, we give an indication of the prevalence of the themes by using quantifiers such as ‘most’, ‘many’, ‘some’, and ‘a few’. Quotes presented here as part of the analyses are followed by a pseudonymised interviewee number. 
Further details on our initial round of coding and the mapping of the initial codes to the main final themes can be found in Appendix~\ref{sec:thematic-analysis}.

\paragraph{Interviewee Demographics.}
The analysed interviews corresponded to a set of interviewees that was relatively balanced in terms of gender and diverse in terms of age and ethnicity (see Table~\ref{tbl:demographics}). 
Gender self-identification included 25 female (57\%), 18 male (41\%), and 1 non-binary (2\%). 
Age ranged from 19 to 72 (mean: 43, standard deviation: 14), with the most prevalent range being 35--44 (32\%). 


Interviewees were asked to self-identify their ethnicities during the interviews, which gave people space to provide rich and nuanced reflections. In Table~\ref{tbl:demographics}, we provide a summary of participants' ethnicities based on the closest match with the high-level groupings used in the England and Wales census and the Scottish census 2021. Matched ethnicities according to census groupings included 27 `Group A' (61\%), 15 `Group B' (34\%), and 2 `Group M' (5\%). Here, `Group A' consists of the `Asian or Asian British' census group in England and Wales, and the `Asian, Scottish Asian or British Asian' census group in Scotland, and includes interviewees identifying as Indian, Pakistani, Bangladeshi, and Chinese. `Group B' matches best with the `Black, Black British, Caribbean or African' census group in England and Wales, and the `African, Scottish African or British African' and `Caribbean or Black' census groups in Scotland, and includes self-identification as Caribbean, African, and other Black backgrounds. `Group M' includes mixed or multiple ethnicities, referring to similarly named census groups in England and Wales and Scotland. 

We have opted not to provide a more detailed breakdown of ethnicity for two reasons: (a)~nuanced self-identifications of ethnicity are in some cases difficult, and problematic, to map to rigid pre-defined census categories, as was the case for many interviewees in `Group B', specifically between African and Caribbean subcategories; and (b)~as our analysis shows, when asked to disclose their ethnicities, some minoritised ethnic people chose to use broad ethnicity categories as a privacy-preserving behaviour, so our interviews and reporting respect such choices. 
Note these groupings are only provided for summarising the ethnicities of the sample, and should not be misinterpreted as analytical categories, as such practice could lead to homogenisation of people, cultures or heritage. 

Although the interviewees were given the option to be interviewed in their preferred language, only 2 (5\%) out of the 44 requested to be interviewed in languages other than English. This indicates that the overwhelming majority of our selected interviewees were relatively confident in their spoken English proficiency. 
\begin{table}
  \centering
  \caption{Interviewee Age, Gender and Ethnicity Distributions}
  \begin{tabular}{l@{}r}
      \toprule
      Age & Count \\
      \midrule
      18--24 & 5 \\
      25--34 & 8 \\ 
      35--44 & 14 \\ 
      45--54 & 9 \\ 
      55--64 & 2 \\ 
      $\geq$ 65 & 6 \\ 
      \bottomrule
  \end{tabular} 
  \ 
  \begin{tabular}{l@{}r}
      \toprule
      Gender & Count \\
      \midrule
      Female & 25 \\
      Male & 18 \\ 
      Non-binary & 1 \\ 
      \bottomrule
      \\ \\ \\ 
  \end{tabular}
  \ 
  \begin{tabular}{l@{\ \ }r}
      \toprule
      Ethnicity & Count \\
      \midrule
      Group A & 27 \\
      Group B & 15 \\ 
      Group M & 2 \\ 
      \bottomrule
      \\ \\ \\ 
  \end{tabular}
  \label{tbl:demographics}
\end{table}

\subsection{Control over Personal Data \& Privacy}

Concerns about privacy emerged as an important theme in the interviews sampled. Concerns often arose in response to questions about negative experiences of using online services and the general quality of essential services in the UK. 

\paragraph{Personal Data Theft \& Misuse.}
Some individuals were specifically concerned about who had access to view, use or share their data and its safety from theft or misuse. This awareness of privacy-related harms was present and immediate in people's minds.  

\begin{quote}
‘It's the privacy of my information. Who can get hold of that information, is it safe?’ (2004)  
\end{quote}

\begin{quote}
‘That’s the only one thing I have online, the protection of your data. When you leave your information, you’re just thinking, because you hear a lot of stories, people just stealing people’s identity. That’s why a lot of the time I’m trying to think, “shall I do it? shall I not do it?”...that’s the one thing that I don’t like!’ (1050)
\end{quote}

\paragraph{Discomfort with Technology.}
Worries about data misuse also created concerns about doing the `wrong' thing online and self-compromising personal privacy. This was often felt among interviewees who were limited users of the internet and essential online services. This fear in turn influenced how much they engaged with online services from the outset. 

\begin{quote}
‘I don’t like those cookie settings all the time. It makes me feel anxious about whether I might have pressed the wrong thing and how they will use my information if I press the wrong thing.’ (3023)  
\end{quote}

\begin{quote}
‘I feel nervous about my information. It makes me feel anxious thinking about privacy and what people might be doing with my information or what kind of things people will find out and share with others. I think that’s what limits me from using online stuff more.’ (4016, \emph{interview notes}) 
\end{quote}

These experiences highlight an awareness and concern for personal security and privacy protection. Many participants in the sample highlighted privacy as their most significant online concern, beyond usability, accessibility, and access to devices as typically thought of in some previous research (see, e.g., \cite{islam2024digitalised} regarding health, and \cite{hasan2023minority} regarding social housing). However, it is worth noting that control over personal data and privacy have been observed as the most significant concerns of the wider population in other contexts, e.g., in smart homes~\cite{park2023nobody}. It should, therefore, be noted that privacy-related harms were found in some instances to be more immediate to individuals than accessibility or access to devices.   

\paragraph{Concerns about Identity Disclosure.}
Privacy-related concerns subsequently guided people's responses and online behaviours, including questioning motives, rejecting services, or selectively sharing or modifying the information they provide essential services with.   

\begin{quote}
‘I can understand with banks people needing your date of birth, and the GP, hospitals. That's fine, but other than that, I don't see why anyone else needs your date of birth, unless they've got their own reason to do something like a scam.’ (1049) 
\end{quote}

\begin{quote}
‘For a very long time, I didn't do certain things as somebody living in Scotland, which I should have done. I was a bit hesitant because I don't know to what extent that information will be protected. It really held me back in not wanting to use my full surname, right. It held me back, and in some cases I wasn't using my full surname... I’d rather not, because sometimes you don't know where these things lead to, if you know what I mean.’ (3016)  
\end{quote}

\paragraph{Concerns about (Mis)identification with Targeted Groups.}
While some refused to provide any information, opting for ‘prefer not to say’ where possible, others replied to questions inconsistently, particularly when asked for their ethnicity. This ‘inconsistency’, for example, often meant minoritised ethnic people chose to identify under the broadest or obscurest ethnicity categorisation available, e.g., Black other, rather than Black Caribbean, with many intentionally keen to emphasise the ‘British’ or ‘Scottish’ aspect of their identity foremost.  

\begin{quote}
‘I feel like, at any time I am given the option to, “prefer not to say”, I kind of tend to just go for that... I identify myself as probably British, but then I know that question is not really associated in terms of how I associate, it's more in terms of my ethnic background and stuff like that.’ (2006)  
\end{quote}

\begin{quote}
‘I say, “I'm British Pakistani.” I feel that because I was born and raised in this country. I'm British first, then Pakistani. As soon as you put, “Pakistani,” from there it’s like, “Right, okay, mate. Your chances are being brought down.” I've noticed that in Scotland.’ (3002)  
\end{quote}

\begin{quote}
‘I would say, “African Caribbean.” If they say, “What nationality are you?” Then, I would say, “British.” It's such a mess, isn't it? We can't roll the clock back 4000 years or 400 years. We're now doing DNA tests and everybody is trying to identify. That’s the mess the West creates. Some people say they’re Black British. Some say they’re just British.’ (2025)   
\end{quote}

\paragraph{Concerns about Data Sharing Practices.}
The apprehension individuals felt was often because it was unknown, or unknowable, how services might use their personal information both now and in the future, and the impact this may have on decision-making and access to services, credit, affordable energy tariffs, employment or in some cases, their rights as a citizen – as expressed by an individual working within the community:

\begin{quote}
‘Migrants that come in that are going through the asylum process or the refugee process, they're too scared to say anything. They just want to keep their heads down. Not because they're illegal or anything, but even if they're illegal they still have rights too.’ (2007)  
\end{quote}

A series of policies, party politics, systemic racism, austerity, and media rhetoric have contributed to a ‘hostile environment’ in the UK, which has sought to ‘close’ borders to migration and make access to visas, housing, energy, and healthcare services increasingly difficult for newly arrived~\cite{burbidge2024hostile}. The NHS, for example, was once required to share patient demographic data with the Home Office~\cite{worthing2021patients}, while the Windrush Scandal saw the state discriminating against Black British people with deportations and barring access to essential services~\cite{commons202windrush}. These policies have further entrenched a racialised citizenship, impacting British/Scottish nationals of a minoritised ethnic background, as well as those more recently arrived or claiming asylum in the UK.   

Some minoritised ethnic people were, therefore, keenly aware of the impact of systemic racism and prejudice. In some cases, individuals chose to withhold, obscure, or actively reject the extent of information disclosed to services, especially in relation to ethnicity, while some chose not to use online services as a privacy-protective behaviour. This is an important context of fear, anxiety, and lack of trust affecting behaviours in response to the potential repercussions of being ‘identifiable’ and othered. A similar ``online privacy divide'' between ethnic groups have been also observed in social media privacy management behaviours in the US~\cite{wang2024online}.

\paragraph{Summary.}
Minoritised ethnic people who are worried about privacy have \emph{a heightened level of concern} about their personal data being stolen or misused. These concerns lead to an apparently wider adoption of common privacy behaviours, e.g., \emph{selective disclosure}, and more creative ones, e.g., \emph{inconsistent disclosure} of personal data. Negative experiences of online services and the hostile policy environment affect and shape these heightened concerns. 

\subsection{Disclosure Dilemma}

The conscious awareness of privacy-related harms amongst many minoritised ethnic participants led to a high level of intention and planning in controlling information disclosure. 

\paragraph{Ethnicity Disclosure Concerns.}
Some participants discussed specific concerns about the data collection practices of service providers, questioning why certain information is collected and how it is used - leading to conflicting feelings about when and how much information to disclose. The most apparent tensions were related to services asking for ethnicity or information that could infer ‘ethnicity.’  

\begin{quote}
‘I avoid sharing ethnicity information when I can, I’m not 100\% confident about where my information will be used and shared and if it will be taken in a positive perspective. I just say that I’m Black African Caribbean and Black other so it’s a bit purposefully vague.’ (2014, \emph{interview notes}) 
\end{quote}

\begin{quote}
‘If I'm applying for a credit card or something like that... why do you need to know the colour of my skin?’ (2006). 
\end{quote}

\paragraph{Concern about Being Seen as Evasive.}
Due to heightened concerns, some participants subsequently discussed controlling the amount of personal data they provide, such as ethnicity (or other information that could be a proxy for ethnicity, e.g., names), date of birth, address, and age. However, despite fears, some participants felt limited in their choice, wondering if they may be seen as ‘evasive’ and if a privacy-protective behaviour of non-disclosure could negatively impact them.  

\begin{quote}
‘Sometimes people have this feeling that maybe it's [surnames] used to discriminate and that's why some people feel very strongly about not giving that kind of information. Then at the same time you feel, if I don't fill that form, will they think I'm trying to hide something?’ (2004)  
\end{quote}

\paragraph{Ethnicity Proxy Disclosure Concerns.}
Participants who expressed the deepest privacy concerns also discussed experiences of direct racism and systemic racism in essential services including healthcare, e.g., higher rate of maternity deaths in the UK for Black women. The lived experience of adverse racialisation and the potential for discrimination meant some individuals were less or not open to providing personal information.  

\begin{quote}
‘Because I've been asked my name on many an occasion, I've been searched... I've been asked to do all sorts of stuff on many an occasion that is a breach of my privacy. A personal breach of my privacy. So, I'm reluctant to just give my name out willy-nilly or my details out willy-nilly. Even signing that document [participant consent form] that you've given me I was thinking twice about at one point, but I thought no, it's for the greater good so it's fine.’ (3020)  
\end{quote}

\paragraph{Disclosure for the `Greater Good'.}
This concept of the `greater good' in the above quote, however, is a typical example of the cost-benefit analysis participants expressed when discussing the tension between the risk of personal harm versus the potential benefit of organisations having ‘group’ information, in order to improve services. For participant 3020, they wavered yet agreed to sign a university document, even despite being stopped by police services over 40 times. This consideration, or dilemma, about whether to provide information or not, emerged through other interviews:  

\begin{quote}
‘I could understand if for example you find that in diabetes there are more people from the African, South Asian descent who are likely to suffer diabetes, I would understand that. But there are some times whereby they are asking you of your ethnicity, and you're thinking, why do you need that information anyway?’ (2004)  
\end{quote}

Some participants were indifferent to the matter overall, while some actively supported providing information to services, so service providers and policymakers can better understand population needs and monitor access. Participant 2007 noted in their interview that it was the availability of such data in England that exposed the higher rate of COVID-19 related deaths amongst some minoritised ethnic groups. However, one participant expressed that, in practice, providing ethnicity information would have little impact in terms of equality monitoring or accountability. 

\begin{quote}
‘I don't think it makes a difference. I don't think you'll get treated any better. If somebody is a racist, they're going to be a racist regardless of your data. It's all a tick-box exercise. I don't like those things without substance, things that don't have a purpose.’ (1045)  
\end{quote}

\paragraph{Concern about Data Collection Purpose Obscurity.}
The cost-benefit dilemma is further complicated by the variable transparency in service providers behaviour, the clarity of communication to users on the purpose of data collection, and the perception of genuine choice users feel they have. Some participants reasoned there could be positives for other minoritised ethnic people if services processed such information responsibly. Some participants were satisfied in providing ‘sensitive’ information, if it could be used to hold services to account or identify gaps in provision. 

\begin{quote}
‘I understand why it has some value to it, but I understand how it can be perceived as a negative thing. I think the value, if it’s used in a positive way, outweighs the negative because if you can point out there’s an issue here within this group of people that they should be aware of, and just let them know. Say, “Look, this is what we’ve found. We think the issue has stemmed from this”... but it should all be about choice really.’ (2010)  
\end{quote}

Disclosing personal information, especially ethnicity, is also at the cost of exposing individuals to privacy-related harms, or alternatively impacting the potential for data-led service improvement. Although this dilemma may be universally experienced by any user, it is specifically problematic for minoritised ethnic people who are already marginalised and experience poorer access to, and outcomes from, essential services. This is in addition to being in a less privileged position to start with: there is more at stake, and more risks involved for minoritised ethnic people, when navigating challenges to protect their privacy and accessing services. Limiting self-disclosure as a privacy-protecting measure has also been observed in other marginalised groups, e.g., undocumented immigrants in the US have reported restricting their posts, especially photos, in social media~\cite{guberek2018keeping}. 

\paragraph{Summary.}
Minoritised ethnic people are particularly concerned about \emph{excessive collection of ethnicity data} by essential online services and \emph{being seen as evasive} in case of non-disclosure. While many minoritised ethnic people are willing to provide such information for the \emph{`greater good'} of improving data-led decision making and services, some are doubtful if such disclosure would make any difference. \emph{Systemic racism} and the widespread \emph{opacity of data collection purpose} are some of the root causes that affect the accentuated concerns about disclosure of ethnicity and create a \emph{high-stakes dilemma} regarding disclosure of such information with which minoritised ethnic people grapple on an everyday basis. 

\subsection{Fear of Repercussions}

The concerns minoritised ethnic people had about privacy-related harms and when/if to disclose personal information were often driven by a fear of repercussions. 

\paragraph{Information Withholding Practices.}
Many were often concerned about potentially doing the ‘wrong’ thing, and jeopardising their access to services or being actively penalised for it. Fears were often a product of poor prior experiences related to systemic racism, austerity, and a hostile environment – all of which has increased conditions of reduced access to services, higher costs, delays or inadequate treatment. Many expressed being fearful of further negative consequences online, such as exclusion, increased charges, or being discriminated against.   

\begin{quote}
‘I’m aware of applications for benefits being sabotaged, small elements in the DWP [Department for Work and Pensions] department that were racist to prevent and delay applications amongst minority ethnic people. So, I’m very aware of how information could be used against you.’ (2014, \emph{interview notes}) 
\end{quote}
 
\begin{quote}
‘They say it's anonymous. It's almost that thing that if they know where you're from, it could easily sway their decision, despite all the reassurances. You get reassurances that it's not going to be given to anyone, but also, it's unnecessary.’ (1045)  
\end{quote}

\paragraph{Protection from Vulnerability \& Distrust.}
Some participants expressed concerns related to a lack of trust about how services may utilise or interpret user information and how securely this information would be held. This distrust, especially towards technology giants, has been observed in other contexts as well~\cite{usman2023distrust}. In response, some people engaged in self-censorship or withdrew altogether from services as a privacy-protective behaviour.   

\begin{quote}
‘I think we're living in very untrustworthy, very cynical times, and all I'm trying to do is just protect myself and my son especially, because he'll be more vulnerable with his disabilities, so that's all I'm trying to do, really.’ (2023)  
\end{quote}

\begin{quote}
‘I didn't want that information to go out everywhere... I have concerns... about how will they misinterpret this - or how will they see this? I think there should be some time limit where they should store it and then get rid of it or something. Maybe at that time, you discuss something personal, but then that data is everywhere, and bearing in mind, other people can access it straightaway.’ (4017)  
\end{quote}

\begin{quote}
‘Interviewer: Earlier on you mentioned that you don't trust services online as much...  

Participant 2004: No, I always feel that if I give my - you hear all sorts of nightmare stories where somebody has stolen!... you've seen how information in the NHS can be compromised. It's one of my things that stops me from doing a lot of stuff online.’  
\end{quote}

\paragraph{The `Spill-over' Effect.}
The stakes of “things going wrong” are heightened for minoritised ethnic people who are burdened by disproportionate risks of harm from socially constructed views, which can feed stereotyping, discrimination, and prejudice against them, when their ethnicity information becomes visible or assumed. This is consistent with many minoritised ethnic people’s lived experience of being racialised and marginalised in everyday life, including often being dismissed, undermined, and distrusted because of ‘everyday racism’, as coined by Essed~\cite{essed1991understanding}.  

While few of the interviewees drew a direct link between their negative experiences of racism on social media and their perceptions of online services, many people were, nonetheless, hyper-aware of how other users may interact with them, if parts of their identity are known publicly.   

\begin{quote}
‘I think it's just a toxic environment and I've had experience of very negative, racist comments and abuse on social media.’ (3020) 
\end{quote}

\begin{quote}
‘Oh, it all happens on social media because everybody is nice to each other face-to-face, but once you're behind that keyboard - I don't know. One of my daughters wears a hijab and the other ones don't, and I still feel very like, oh, my gosh, I've got to make her grow a backbone because I don't know how that's going to affect her.’ (1045) 
\end{quote}

\begin{quote}
‘I used to run an anti-racism page with other people of colour, which I found was quite a harmful space, because people were allowed to be abusive to you. I started my own anti-racism education platform a few months ago, and the way I've set it up is that no one can comment.’ (3004)  
\end{quote}

\begin{quote}
‘Participant 2009: You might find on social media, we have got groups, in our neighbourhood... you just find it isn't easy because with people, it is quite difficult to say it, but sometimes people just look at us and then they are like, you know...  

Interviewer: Racist?   

Participant 2009: Yes, we are like some burden or something. Even though we do experience it, but we just have to live with it, and that is why most of the time, I'm not really into this, posting stuff on Facebook or Instagram, somewhere, no... that may ruin your mental stability.’   
\end{quote}

There is risk of a potential `spill-over' effect from these negative general online experiences, as some participants suggested they felt a lack of trust towards any interaction with services through digital means, because of the harms they experienced elsewhere online, especially on social media. This leaves minoritised ethnic people in a disproportionately more vulnerable position when they navigate privacy challenges, given the historic, systemic, and personal racism many minoritised ethnic people encounter, both online and offline, which cannot be necessarily disentangled from each other.  Many individuals in our study were, therefore, found to be guarding against risks of repercussions by withholding information, avoiding, or refusing to engage with essential services. Negotiating trust and safety is a constant and temporal process.

\paragraph{Summary.}
Minoritised ethnic people have distinct concerns about receiving \emph{discriminatory services} as a result of their ethnicity information being misused. This is amplified by perceptions of \emph{inadequate data protection practices} by essential services. Concerns are affected not only by individual experiences of using essential services, but through negative experiences of other online services, especially social media. This strongly suggests that there is an unfortunate \emph{`spill-over' effect} extending the lack of trust in online services in general, and specifically social media, to essential services. 

\section{Discussion}

The data illustrates the importance of trust, which is embedded across each theme presented in this paper. The impact of events or experiences which may have occurred ‘offline’ are inextricably linked to the concerns and  awareness of security and privacy-related harms. Negative perceptions of trust in services affect how people interact and engage with online services, if at all. On a more systemic level, if users withdraw from services, they may be less likely to hold systems to account or provide the feedback necessary to make systems more equitable. Often, we lack understanding in service design in how to engage with socially constructed issues, which nevertheless interact with and influence security and privacy-related concerns and behaviours. There is also a temporality involved when users consider their privacy needs. These are continually negotiated and influenced by personal experiences, and for minoritised ethnic people, these needs are heightened at different times in response to external events, e.g., data breaches, hostile environment, and racist attacks. 

\subsection{Limitations}
While this paper has focused on users' privacy concerns, there were some interviewees who expressed little concern or were unsure how much concern they should have for security and privacy-related harms. Our wider group of participants, i.e., the 101, were relatively more concerned about the risk of scams, financial fraud or data leakage compared to racial profiling in general. However, it is well-known that data profiling can confer ethnicity and thus increase privacy risks amongst minoritised ethnic users who may be exposed to greater threats because of systemic bias and racism. Lack of awareness in this regard is concerning as research has shown that privacy-protecting behaviours depend on awareness of the need for privacy foremost, besides the usability and usefulness of specific privacy tools~\cite{coopamootoo2020usage}. 

However, this limited concern is reasonable to expect as interviews were focused on individual experiences, as opposed to exploring the processes of digital profiling, which can be challenging even for well-informed individuals to fully comprehend due to the opacity and pervasiveness of profiling across online interactions. Much of the privacy implications of our online activities are unknown, meanwhile the privacy landscape continues to change at pace, meaning existing good practice or knowledge will require periodic re-assessment. 

Moreover, we caution against individualising privacy and security. There should be systemic solutions and responsibilities upheld within services to protect against privacy and security-related harms, e.g., security and privacy by design, auditing, and impact assessment. While attention can be directed towards the knowledge and behaviours of individuals, it is necessary for services to be user-centred and demonstrate they can be entrusted with users' data. 

We attempted to reach a diverse range of minoritised ethnic people based on the largest ethnic groups as labelled by the UK censuses. However, we wish to make clear that these labellings do not represent the complexity of how people identify as social, cultural, and historical contexts influence how people define their ethnicity, if at all. Definitions are also situated within a political context considered ‘hostile’ towards particular minoritised ethnic communities, and one which can essentialise and homogenise a diversity of people who represent global majorities. This nuance and complex interplay between ‘official’ labels, institutionalised by services and personal identification, is discussed further in our analysis, but is expressed by one participant in response to an interview question about defining their ethnicity: 

\begin{quote}
‘It always says either British, Indian, Pakistani or Bangladeshi, it never gives you the Muslim Gujarati... it’s not targeting the small minority of the Gujarati Muslim community... reluctantly, I then have to click Indian, and then where it says 'other' in the box, I've put down, “I'm a British Muslim, Gujarati,” and then I put in brackets, “Indian.” It's a mouthful. Just that my identity... Indian? with the historical issues, honestly, I don't really want to associate, because when you're doing the surveys and the stats for our government or whatever, “Oh, another Indian, yes, they're meeting the targets,”... but our community's underrepresented.’ (4033) 
\end{quote}

We, therefore, acknowledge the limitations of the data in this study because it cannot be considered representative of the diversity of experiences and characteristics of the people who comprise minoritised ethnic populations in the UK. Rather, the purpose of presenting interview data was to offer empirical depth and insight into the privacy-related experiences of minoritised ethnic people, in a context where research on such communities is limited, particularly in the UK.  

\subsection{Risks of Under-reporting}
Our analysis highlights the difficulties many minoritised ethnic individuals encounter when asked for personal information, particularly that related to ethnicity. We recognise the complexity of ethnicity as a data label for its historical, political, and social location, which does not capture the nuances of how people feel and define themselves. There are, of course, well-known risks of stereotypical profiling and bias through proxy discrimination. Yet, if services have inconsistencies across datasets, there are some drawbacks to auditing services and any data-driven methods used to enhance services for fairness. Anti-discrimination models require the inclusion of sensitive data to detect and avoid discriminatory outcomes~\cite{favaretto2019big}. Patchy datasets could, therefore, impede effective anti-discrimination work within services. This means developing trust and communicating privacy terms and data collection practices with users is important and essential to the effective auditing and evaluation of services.

\subsection{Implications for Research}

Here we discuss specific implications for research with minoritised people, including in the field of security and privacy, as informed by our experience conducting this work. 

\paragraph{Facilitate privacy-protecting behaviour.}
The decision by some participants to decline audio recording was often indicative of heightened privacy concerns, and even the direct contact with a university researcher connected to a community organisation known to them was not enough to mitigate concerns or allay fears, as one participant described:  

\begin{quote}
‘She doesn't want to be recorded for this interview because she doesn't want her voice to be heard in other ways. So, for example, she never sent any voice note to WhatsApp. She always types it in, so she is very conservative in that her voice is not going anywhere. In that way, she feels she stays safe.’ (1051, \emph{interview notes, spoken via a translator}).
\end{quote}

This suggests researchers should consider additional privacy-protecting measures when interacting with and collecting data from minoritised people. Best practice in this regard should provide flexibility of consent and data recording methods to accommodate heightened sensitivities towards disclosure of personal data. Such varying security and privacy needs have been observed in other marginalised groups too, e.g., those marginalised based on gender~\cite{coopamootoo2023equal}.

\paragraph{Allow self expression of identity.}
Approaches employed by the participants in this study to express various aspects of their identity and especially their ethnicity were quite varied, context-based, and creative. This was the case both for their direct responses to self-identify their ethnicity and for their broader discussions on how their ethnicity interacted with their security and privacy-related decision making and behaviour. While in certain contexts, some would opt for identification in broad terms, e.g., Asian rather than Pakistani, or emphasise their `Britishness', e.g., British Asian, in other contexts some would prefer a more detailed and nuanced specification, as the following participant elaborated: 

\begin{quote}
‘There was some who say, “Why do you need to know? If I'm black, I'm black. Why do you need to know if I'm Caribbean?” I say, Well, there's a difference between an African and a Caribbean first, and also because of our climate, our culture, our dietary, there's so many things... I think that's important for the system to understand, so that when it comes to the issues, it makes this bit of the system easier to tackle rather than to start from scratch.’ (2007)  
\end{quote}

Hence, we recommend researchers design their data collection methods in such ways that not only permit but also encourage nuanced self-expression of aspects of identity, especially ethnicity, and including gender and religion. This is particularly relevant to those conducting research on security and privacy issues as we found that allowing for an open dialogue in such matters often leads to further insight into privacy-related concerns and behaviours. We advise against rigid and pre-determined sets of answers in this regard, especially on ethnicity, as they neither encourage nuanced discussions nor may provide accurate data. 
Chen et al.~\cite{ChenSRT23why} also recommend self-identification and open-ended questions for collecting ``race'' and ethnicity data from participants. 
Similar recommendations have been discussed in the literature in the context of presentation of sexuality and gender identities~\cite{devito2018too}.

\subsection{Lessons Learned}

We discuss recommendations for \emph{service designers} and \emph{policymakers} based on our analyses in the following. 

\paragraph{Understand the social context.}
Data exists within systems of power, known to replicate the systemic biases of wider society~\cite{miceli2022studying}, and many of our interviewees were concerned about this. We recommend service designers, and those who work directly or indirectly with data throughout its lifecycle, improve their awareness of the socio-cultural context of particular data labels and how this may impact how people subsequently interact with digital services. This awareness should be translated into designs that are sensitive to social contexts and provide choice. A similar recommendation to identify ``cultural assumptions embedded in computer security technologies and account for them in technology designs'' has been echoed in the context of refugees in the United States~\cite{Simko2018refugees}. Given we often cannot ‘necessarily predict when a certain form of information processing will produce predictive privacy harms’, it is necessary and urgent for a greater understanding of privacy and security issues, and to continually interrogate the subjectivities inherent within the use of data and digital systems and how they are experienced by different users~\cite[p.\,106]{crawford2014big}. More concretely, we recommend policymakers actively redress the institutional injustices experienced by minoritised ethnic people (e.g., Windrush Scandal), to help build trust and see such work as a necessary part of digital inclusion strategies and audits or regulations of essential service providers. Furthermore, policymakers should encourage online services such as social media to adopt and effectively implement zero-tolerance policies on racist behaviour on these platforms.

\paragraph{Ease the dilemma.}
Minoritised ethnic people were concerned about violation of their privacy by online services and many of those we interviewed mentioned experiences of scams and online fraud along with their concerns, which strongly suggest these are causally linked – similar links have been shown for other groups (see, e.g., \cite{chen2017securing} and \cite{aldaraani2022online}). Hence, we recommend that service designers minimise data collection, especially regarding ethnicity and proxies thereof, and clearly explain the purpose of any such data collection. Many minoritised ethnic people stated that if they are clearly informed of the underlying reason(s) for data collection, they will be more likely to share data. In cases where there is a clear benefit for minoritised ethnic communities and wider society, we recommend service designers communicate openly and appeal to people's willingness to share data for the 'greater good'. However, in cases where data sharing is optional, service designers should make it clear that there will be no negative consequences for non-disclosure to allay any concerns. Finally, service designer should recognise the right to exercise agency over control of personal data and allow for various levels of disclosure of ethnicity to accommodate selective disclosure. These practices will help ease the dilemma faced by minoritised ethnic people on an everyday basis. 

\paragraph{Build trust.}
Trust, or lack thereof, was found to be a salient consideration for minoritised ethnic people in deciding whether to engage or how to interact with online services. Negative experiences on the wider web, including experiences of racism, online tracking, and identity theft, were frequently mentioned as contributing factors to lack of trust in online services in general. Therefore, we recommend that designers of essential services make every attempt to distinguish these services from other online services to reduce this `spill-over' effect by taking a systematic approach to creating usable, accessible, and privacy-preserving services that avoid harmful `dark patterns' prevalent in other online services. By adopting and communicating strong data protection practices, service designers can help alleviate concerns about data theft and misuse. The literature provides similar recommendations for password manager designers to adopt privacy labels communicating incorporated security mechanisms~\cite{emami2021privacy, li2022understanding, klivan2023would}.

\section{Conclusion}
This work has examined the intersection of data privacy, security, and the lived experiences of minoritised ethnic people in the UK, with a particular focus on essential online services in the health, social housing, and energy sectors. Through qualitative analysis of interviews with 44 participants, our work has provided valuable insight into how privacy and security concerns are shaped by experiences of racism and marginalisation, which subsequently influences engagement with online services. Our findings highlight not only widespread fear of digital discrimination and potential misuse of personal data, but also a level of ambivalence among some individuals who feel disconnected due to limited awareness of how their data is utilised or belief in the positive potential of data sharing.  

We have focused on minoritised ethnic people, a group currently underrepresented in privacy and security research, who often face specific barriers, such as trust issues arising from past experiences of racism, and how this social context influences online service engagement. We offer evidence of how privacy and security concerns may manifest in these communities, through an interdisciplinary approach. Integrating anti-racism with privacy research, has enabled a nuanced and contextually grounded analysis, shedding light on the complex relationship between racialisation, marginalisation, and interactions with essential online services.  

Future research may expand on these findings by investigating how the cultural and historical context of specific minoritised communities influences attitudes and behaviours towards online services. Additionally, further work is needed to understand how systemic racism and ‘offline’ interactions shape perceptions and engagement with online services, e.g., smart meters and online health apps. Our investigation also suggests there may be a significant relationship between people’s experiences of social media and their concerns using essential online services, and call for further exploration of this relationship. Finally, we recommend future work considers how the design of online services can be made more inclusive and trustworthy, considering the diverse needs and experiences of minoritised ethnic populations, to promote digital equity. We believe privacy-enhancing technologies (PETs), as well as policies and service designs that are marginalisation-aware and redress injustice, can play a significant role towards achieving this goal. Co-designing such technologies and policies with marginalised users is one of the research directions we plan to pursue in this regard.

\section*{Acknowledgements}
We acknowledge the significant contribution of the data collection team and the care with which they approached the interviews. The team included Sara Bailey, Sacha Hasan, Farjana Islam, and Aunam Quyoum. 
We wish to thank all the interview participants and third sector/community partner organisations, who shared their time and expertise generously and contributed significantly to the project and findings underpinning this paper. 
We would like to acknowledge Gina Netto and Laura Whyte for their leadership and management of the wider project. 
The constructive comments from the anonymous reviewers of SOUPS~2025 are also appreciated. 
This work was supported by the UKRI Strategic Priorities Fund under Grants 
EP/W03235X/1 
and 
EP/W032058/1. 



\bibliographystyle{plain}
\bibliography{privacy-paper-refs}

\begin{thebibliography}{10}

\bibitem{ahmed2017digital}
Syed~Ishtiaque Ahmed, Md~Romael Haque, Jay Chen, and Nicola Dell.
\newblock Digital privacy challenges with shared mobile phone use in {Bangladesh}.
\newblock {\em Proceedings of the ACM on Human-computer Interaction}, 1(CSCW):1--20, 2017.

\bibitem{ahmed2015privacy}
Tousif Ahmed, Roberto Hoyle, Kay Connelly, David Crandall, and Apu Kapadia.
\newblock Privacy concerns and behaviors of people with visual impairments.
\newblock In {\em Proceedings of the 33rd Annual ACM Conference on Human Factors in Computing Systems}, pages 3523--3532, 2015.

\bibitem{aldaraani2022online}
Najla Aldaraani, Helen Petrie, and Siamak~F Shahandashti.
\newblock Online security attack experience and worries of young adults in the {United Kingdom}.
\newblock In {\em International Symposium on Human Aspects of Information Security and Assurance}, pages 300--309. Springer, 2022.

\bibitem{anthony2017toward}
Denise Anthony, Celeste Campos-Castillo, and Christine Horne.
\newblock Toward a sociology of privacy.
\newblock {\em Annual Review of Sociology}, 43(1):249--269, 2017.

\bibitem{atienza2015consumer}
Audie~A Atienza, Christina Zarcadoolas, Wendy Vaughon, Penelope Hughes, Vaishali Patel, Wen-Ying~Sylvia Chou, and Joy Pritts.
\newblock Consumer attitudes and perceptions on mhealth privacy and security: findings from a mixed-methods study.
\newblock {\em Journal of Health Communication}, 20(6):673--679, 2015.

\bibitem{braun2021thematic}
Virginia Braun and Victoria Clarke.
\newblock {\em Thematic analysis: A practical guide}.
\newblock SAGE publications Ltd, 2021.

\bibitem{braun2023toward}
Virginia Braun and Victoria Clarke.
\newblock Toward good practice in thematic analysis: Avoiding common problems and be(com)ing a knowing researcher.
\newblock {\em International Journal of Transgender Health}, 24(1):1--6, 2023.

\bibitem{burbidge2024hostile}
Manon Burbidge, Stefan Bouzarovski, Karen Lucas, and Saskia Warren.
\newblock Hostile environments: Housing and asylum policies as drivers of energy deprivation among {UK} refugee communities.
\newblock {\em Housing, Theory and Society}, pages 1--17, 2024.

\bibitem{chen2017securing}
Hongliang Chen, Christopher~E Beaudoin, and Traci Hong.
\newblock Securing online privacy: An empirical test on internet scam victimization, online privacy concerns, and privacy protection behaviors.
\newblock {\em Computers in Human Behavior}, 70:291--302, 2017.

\bibitem{ChenSRT23why}
Yiqun~T. Chen, Angela D.~R. Smith, Katharina Reinecke, and Alexandra To.
\newblock Why, when, and from whom: considerations for collecting and reporting race and ethnicity data in {HCI}.
\newblock In {\em {CHI}}, pages 395:1--395:15. {ACM}, 2023.

\bibitem{coopamootoo2020usage}
Kovila~PL Coopamootoo.
\newblock Usage patterns of privacy-enhancing technologies.
\newblock In {\em Proceedings of the 2020 ACM SIGSAC Conference on Computer and Communications Security}, pages 1371--1390, 2020.

\bibitem{coopamootoo2023equal}
Kovila~PL Coopamootoo and Magdalene Ng.
\newblock {``Un-Equal Online Safety?''} {A} gender analysis of security and privacy protection advice and behaviour patterns.
\newblock In {\em 32nd USENIX Security Symposium}, pages 5611--5628, 2023.

\bibitem{cornejo2016vulnerability}
Raymundo Cornejo, Robin Brewer, Caroline Edasis, and Anne~Marie Piper.
\newblock Vulnerability, sharing, and privacy: Analyzing art therapy for older adults with dementia.
\newblock In {\em Proceedings of the 19th ACM Conference on Computer-Supported Cooperative Work \& Social Computing}, pages 1572--1583, 2016.

\bibitem{crawford2014big}
Kate Crawford and Jason Schultz.
\newblock Big data and due process: Toward a framework to redress predictive privacy harms.
\newblock {\em Boston College Law Review}, 55:93, 2014.

\bibitem{dencik2018data}
Lina Dencik, Arne Hintz, Joanna Redden, and Harry Warne.
\newblock Data scores as governance: Investigating uses of citizen scoring in public services project report.
\newblock Cardiff University: Open Society Foundations, 2018.
\newblock \url{https://orca.cardiff.ac.uk/id/eprint/117517}.

\bibitem{devito2018too}
Michael~A DeVito, Ashley~Marie Walker, and Jeremy Birnholtz.
\newblock {`Too Gay for Facebook'}: Presenting {LGBTQ+} identity throughout the personal social media ecosystem.
\newblock {\em Proceedings of the ACM on Human-Computer Interaction}, 2(CSCW):1--23, 2018.

\bibitem{dodel2023inequalities}
Mat{\'\i}as Dodel.
\newblock Inequalities and privacy in the context of social media.
\newblock In {\em The Routledge handbook of privacy and social media}, pages 204--214. Routledge, 2023.

\bibitem{dosono2015m}
Bryan Dosono, Jordan Hayes, and Yang Wang.
\newblock {“I’m Stuck!”}: A contextual inquiry of people with visual impairments in authentication.
\newblock In {\em Eleventh Symposium On Usable Privacy and Security (SOUPS 2015)}, pages 151--168, 2015.

\bibitem{emami2021privacy}
Pardis Emami-Naeini, Janarth Dheenadhayalan, Yuvraj Agarwal, and Lorrie~Faith Cranor.
\newblock Which privacy and security attributes most impact consumers’ risk perception and willingness to purchase iot devices?
\newblock In {\em 2021 IEEE Symposium on Security and Privacy (SP)}, pages 519--536. IEEE, 2021.

\bibitem{essed1991understanding}
Philomena Essed.
\newblock {\em Understanding everyday racism: An interdisciplinary theory}, volume~2.
\newblock Sage, 1991.

\bibitem{favaretto2019big}
Maddalena Favaretto, Eva De~Clercq, and Bernice~Simone Elger.
\newblock Big data and discrimination: perils, promises and solutions. a systematic review.
\newblock {\em Journal of Big Data}, 6(1):1--27, 2019.

\bibitem{forte2017privacy}
Andrea Forte, Nazanin Andalibi, and Rachel Greenstadt.
\newblock Privacy, anonymity, and perceived risk in open collaboration: A study of {Tor} users and {Wikipedians}.
\newblock In {\em Proceedings of the 2017 ACM Conference on Computer Supported Cooperative Work and Social Computing}, pages 1800--1811, 2017.

\bibitem{fullenwieder2018settler}
Lara Fullenwieder and Adam Molnar.
\newblock Settler governance and privacy: {Canada}’s {Indian Residential School Settlement Agreement} and the mediation of state-based violence.
\newblock {\em International Journal of Communication}, 12:1332--1349, 2018.

\bibitem{gangadharan2017downside}
Seeta~Pe{\~n}a Gangadharan.
\newblock The downside of digital inclusion: Expectations and experiences of privacy and surveillance among marginal internet users.
\newblock {\em New Media \& Society}, 19(4):597--615, 2017.

\bibitem{guberek2018keeping}
Tamy Guberek, Allison McDonald, Sylvia Simioni, Abraham~H Mhaidli, Kentaro Toyama, and Florian Schaub.
\newblock Keeping a low profile? technology, risk and privacy among undocumented immigrants.
\newblock In {\em Proceedings of the 2018 CHI Conference on Human Factors in Computing Systems}, pages 1--15, 2018.

\bibitem{hasan2023minority}
Sacha Hasan and Yingfang Yuan.
\newblock Minority ethnic vulnerabilities in the use of digital housing services across age groups.
\newblock In {\em The European Network for Housing Research (ENHR) Conference 2023}. The European Network for Housing Research (ENHR), 2023.

\bibitem{hayes2019cooperative}
Jordan Hayes, Smirity Kaushik, Charlotte~Emily Price, and Yang Wang.
\newblock Cooperative privacy and security: Learning from people with visual impairments and their allies.
\newblock In {\em Fifteenth Symposium on Usable Privacy and Security (SOUPS 2019)}, pages 1--20, 2019.

\bibitem{islam2024digitalised}
Farjana Islam, Sara Bailey, and Gina Netto.
\newblock Digitalised primary care in the {UK}: a qualitative study of the experiences of minoritised ethnic communities.
\newblock {\em The British Journal of General Practice}, 74(749):e823, 2024.

\bibitem{rowntree2025poverty}
{Joseph Rowntree Foundation}.
\newblock {UK Poverty 2025}: The essential guide to understanding poverty in the {UK}.
\newblock Report, 2025.
\newblock \url{https://www.jrf.org.uk/uk-poverty-2025-the-essential-guide-to-understanding-poverty-in-the-uk}.

\bibitem{kam2020latinx}
Jennifer~A Kam, Andy~J Merolla, and Andrew~C High.
\newblock Latinx immigrant youth’s indirect and direct disclosures about their family-undocumented experiences, received emotional support, and depressive symptoms.
\newblock {\em Communication Research}, 47(4):599--622, 2020.

\bibitem{klivan2023would}
Sabrina Klivan, Sandra H{\"o}ltervennhoff, Nicolas Huaman, Yasemin Acar, and Sascha Fahl.
\newblock ``{Would} you give the same priority to the bank and a game? {I} do not!'' {Exploring} credential management strategies and obstacles during password manager setup.
\newblock In {\em Nineteenth Symposium on Usable Privacy and Security (SOUPS 2023)}, pages 171--190, 2023.

\bibitem{lerner2020privacy}
Ada Lerner, Helen~Yuxun He, Anna Kawakami, Silvia~Catherine Zeamer, and Roberto Hoyle.
\newblock Privacy and activism in the transgender community.
\newblock In {\em Proceedings of the 2020 CHI Conference on Human Factors in Computing Systems}, pages 1--13, 2020.

\bibitem{li2022understanding}
Yucheng Li, Deyuan Chen, Tianshi Li, Yuvraj Agarwal, Lorrie~Faith Cranor, and Jason~I Hong.
\newblock Understanding ios privacy nutrition labels: An exploratory large-scale analysis of app store data.
\newblock In {\em CHI Conference on Human Factors in Computing Systems Extended Abstracts}, pages 1--7, 2022.

\bibitem{lyu2024understanding}
Tu~Lyu, Yulin Guo, and Hao Chen.
\newblock Understanding the privacy protection disengagement behaviour of contactless digital service users: The roles of privacy fatigue and privacy literacy.
\newblock {\em Behaviour \& Information Technology}, 43(10):2007--2023, 2024.

\bibitem{madden2017privacy}
Mary Madden.
\newblock Privacy, security, and digital inequality.
\newblock Data \& Society Research Institute, Report, 2017.
\newblock \url{https://datasociety.net/library/privacy-security-and-digital-inequality}.

\bibitem{maris2020tracking}
Elena Maris, Timothy Libert, and Jennifer~R Henrichsen.
\newblock Tracking sex: The implications of widespread sexual data leakage and tracking on porn websites.
\newblock {\em New Media \& Society}, 22(11):2018--2038, 2020.

\bibitem{masur2025comparative}
Philipp~K Masur, Dmitry Epstein, Kelly Quinn, Carsten Wilhelm, Lemi Baruh, and Christoph Lutz.
\newblock Comparative privacy research: Literature review, framework, and research agenda.
\newblock {\em The Information Society}, pages 1--22, 2025.

\bibitem{mentis2020illusion}
Helena~M Mentis, Galina Madjaroff, Aaron Massey, and Zoya Trendafilova.
\newblock The illusion of choice in discussing cybersecurity safeguards between older adults with mild cognitive impairment and their caregivers.
\newblock {\em Proceedings of the ACM on Human-Computer Interaction}, 4(CSCW2):1--19, 2020.

\bibitem{miceli2022studying}
Milagros Miceli, Julian Posada, and Tianling Yang.
\newblock Studying up machine learning data: Why talk about bias when we mean power?
\newblock {\em Proceedings of the ACM on Human-Computer Interaction}, 6(GROUP):1--14, 2022.

\bibitem{murji2017racism}
Karim Murji.
\newblock {\em Racism, policy and politics}.
\newblock Policy Press, 2017.

\bibitem{park2023nobody}
Sunyup Park, Anna Lenhart, Michael Zimmer, and Jessica Vitak.
\newblock ``{Nobody}'s happy'': Design insights from privacy-conscious smart home power users on enhancing data transparency, visibility, and control.
\newblock In {\em {SOUPS}}. {USENIX} Association, 2023.

\bibitem{prime-technical-toolkit}
{PRIME Technical Toolkit}.
\newblock Protecting Minority Ethnic Communites Online (PRIME) Project, Technical Toolkit.
\newblock \url{https://primetoolkit.co.uk/}.

\bibitem{quyoum2024valuing}
Aunam Quyoum and Mark Wong.
\newblock Valuing lived experience and co-design solutions to counter racial inequality in data and algorithmic systems in {UK}’s digital services.
\newblock {\em Information, Communication \& Society}, pages 1--17, 2024.

\bibitem{Reichel2017race}
Matt Reichel.
\newblock Race, class, and privacy: A critical historical review.
\newblock {\em International Journal of Communication}, 11(0), 2017.

\bibitem{Rennie2018privacy}
Ellie Rennie, Tyson Yunkaporta, and Indigo Holcombe-James.
\newblock Privacy versus relatedness: managing device use in {Australia}'s remote {Aboriginal} communities.
\newblock {\em International Journal of Communication [online]}, pages 1291--1310, 2018.

\bibitem{Sannon2022privacy}
Shruti Sannon and Andrea Forte.
\newblock Privacy research with marginalized groups: What we know, what's needed, and what's next.
\newblock {\em Proceedings of the ACM on Human-Computer Interaction}, 6({CSCW2}):1--33, 2022.

\bibitem{Simko2018refugees}
Lucy Simko, Ada Lerner, Samia Ibtasam, Franziska Roesner, and Tadayoshi Kohno.
\newblock Computer security and privacy for refugees in the {United States}.
\newblock In {\em {IEEE} Symposium on Security and Privacy}, pages 409--423. {IEEE} Computer Society, 2018.

\bibitem{solove2006taxonomy}
Daniel~J Solove.
\newblock A taxonomy of privacy.
\newblock {\em University of Pennsylvania Law Review}, 154(3):477, 2006.

\bibitem{srinivasan2018privacy}
Janaki Srinivasan, Savita Bailur, Emrys Schoemaker, and Sarita Seshagiri.
\newblock The poverty of privacy: Understanding privacy trade-offs from identity infrastructure users in {India}.
\newblock {\em International Journal of Communication}, 12:1228--1247, 2018.

\bibitem{usman2023distrust}
Warda Usman, Jackie Hu, McKynlee Wilson, and Daniel Zappala.
\newblock Distrust of big tech and a desire for privacy: Understanding the motivations of people who have voluntarily adopted secure email.
\newblock In {\em Nineteenth Symposium on Usable Privacy and Security (SOUPS 2023)}, pages 473--490, 2023.

\bibitem{vanstaden2024localised}
Stefanus Van~Staden and Nicola~J Bidwell.
\newblock Localised trust in a globalised knot: Designing information privacy for {Digital-ID}.
\newblock {\em ACM Journal on Computing and Sustainable Societies}, 2(1):1--37, 2024.

\bibitem{walker2019moving}
Ashley~Marie Walker, Yaxing Yao, Christine Geeng, Roberto Hoyle, and Pamela Wisniewski.
\newblock Moving beyond 'one size fits all' research considerations for working with vulnerable populations.
\newblock {\em Interactions}, 26(6):34--39, 2019.

\bibitem{wang2024online}
Laurent~H Wang and Miriam~J Metzger.
\newblock The online privacy divide: testing resource and identity explanations for racial/ethnic differences in privacy concerns and privacy management behaviors on social media.
\newblock {\em Communication Research}, 2024.

\bibitem{wang2018inclusive}
Yang Wang.
\newblock Inclusive security and privacy.
\newblock {\em IEEE Security \& Privacy}, 16(4):82--87, 2018.

\bibitem{commons202windrush}
Wendy Williams.
\newblock {Windrush Lessons Learned Review}.
\newblock Report, House of Commons, UK Government, 2020.

\bibitem{wong2024minoritised}
Mark Wong, Aunam Quyoum, and Ankita Mishra.
\newblock Minoritised ethnic people’s code of practice for equitable digital services.
\newblock Protecting Minority Ethnic Communites Online (PRIME) Project, Report, 2024.
\newblock \url{https://www.primecommunities.online/code-of-practice}.

\bibitem{worthing2021patients}
Kitty Worthing, Marta~Mojarrieta Galaso, Johanna~Kellett Wright, and Jessica Potter.
\newblock Patients or passports? the ‘hostile environment’in the {NHS}.
\newblock {\em Future Healthcare Journal}, 8(1):28--30, 2021.

\end{thebibliography}

\appendix
\section{Interview Guide (Simplified to Topics)}
\label{sec:interview-questions}

Introductions and Consent 

\noindent 
Interviewee’s Characteristics   
\begin{itemize}
    \item Questions on gender, age, ethnicity, religion, English proficiency, education, occupation, and accommodation status 
\end{itemize}

\noindent 
Online Experience   
\begin{itemize}
    \item Do you personally use the internet? 
    \item Does your household have access to the internet at home? 
    \item What device do you usually access the internet with? 
    \item Which devices do you use and how confidently can you use it? 
    \item What tasks do you struggle to do or need help with? 
    \item Why do you access the internet? 
\end{itemize}
  
\noindent
Online Services – Health   
\begin{itemize}
    \item Have you used internet to access online services for your health?  
    \item Does your GP service have an app? 
    \item Does someone help you to access GP services? How do they offer help? How easy is it to get this help? 
\end{itemize}

\noindent
Online Services – Social Housing   
\begin{itemize}
    \item If relevant, does someone help you to contact your social landlord or apply for social housing? 
    \item If yes or sometimes, who do you normally get help from for contacting your social landlord or applying for social housing? How easy is it to get this help? 
    \item What kind of help do you need to contact your social landlord or to apply for social housing? 
\end{itemize}
  
\noindent
Online Services – Energy   
\begin{itemize}
    \item Do you know who your energy supplier is? Were they the provider when you moved in, or did you change the supplier at any point? 
    \item Do you know that you can change your energy supplier?  
    \item By looking at tariffs different suppliers offer, you may be able to reduce your energy bills. Is this something you will be interested in? 
    \item Do you use a smart meter?  
    \item Would you be interested in the use of a smart meter to help you monitor your energy consumption? 
    \item Do you need any help with accessing energy supply services? If yes, what kind of help do you need? 
    \item Who do you normally get help from for using energy supply services? How easy is it to get this help? 
\end{itemize}
  
\noindent
Online Safety   
\begin{itemize}
    \item Do you take any measures to stay safe online? If yes, what kind of measures do you take? Is there any information you would like to stay safe? 
    \item Have you had any negative experiences while you using online services? 
    \item Has your experience of online harms altered how you use online services? If yes, in what way? 
    \item Have you taken any action in response to online harms? 
\end{itemize}

\noindent
Closing  
\begin{itemize}
    \item How helpful was this interview? 
    \item Is there anything we haven't covered yet as part of your experience of using online services that you would like to talk about? 
\end{itemize}

\section{Community Organisation Partners}
\label{sec:community-org-partners}

Bangladeshi Enterprise and Advisory Project (BEAP) Community Partnership: BEAP provides services directly and through its partner organisations through which it aims to tackle entrenched deprivation and disadvantage in the Bradford area, particularly among the Bangladeshi community, supporting deprived communities to formulate and put into practice their own agendas for positive social change.  

The Caribbean and African Health Network (CAHN): Based in Manchester, CAHN is a Black-led organisation which was set up to tackle the social determinants of ill-health among Caribbean and African communities in the UK. CAHN works with Black communities and cross-sectoral organisations throughout the UK to build community resilience, relationships and a social movement to improve health outcomes for Black people.  

Council of Ethnic Minority Voluntary Organisations (CEMVO Scotland): CEMVO is a national intermediary organisation and strategic partner of the Scottish Government’s Equality Unit. Based in Glasgow, CEMVO aims to build the capacity and sustainability of Scotland’s ethnic minority voluntary sector and the communities the voluntary sector works with.  

The East London Mosque (ELM) \& London Muslim Centre (LMC): The ELM \& LMC is an active Islamic institution based in the borough of Tower Hamlets in London. In addition to daily prayers, the ELM \& LMC delivers advice and counselling, child and adult education, and health awareness programmes to the local community. 

\section{Thematic Analysis Materials}
\label{sec:thematic-analysis}

\begin{figure*}[p]
    \centering
    \caption{Thematic mapping}
    \label{fig:thematic-map}
    \includegraphics[width=\textwidth]{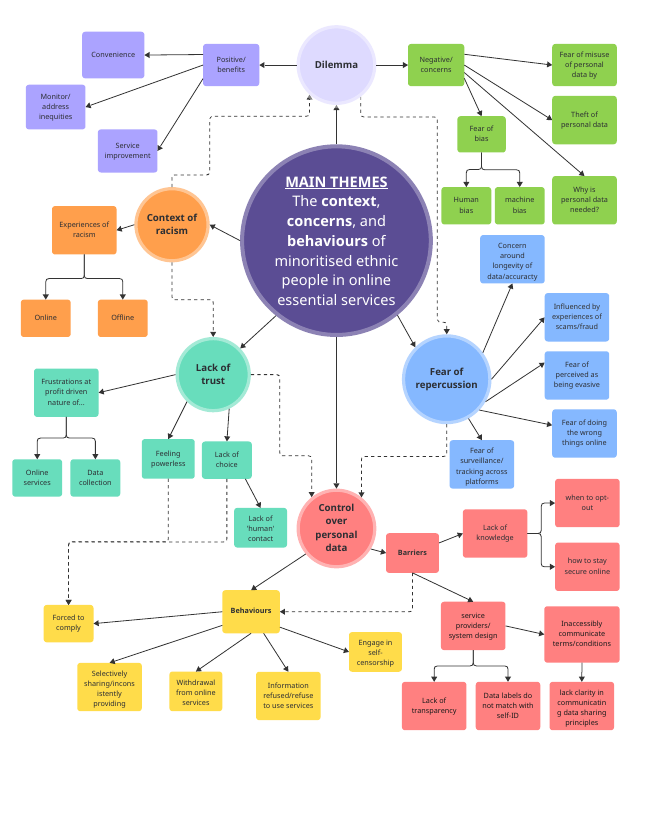}
\end{figure*}

We include further materials on our thematic analysis here. Table~\ref{tbl:themes} shows a summary of the main initial themes and sub-themes identified through the first round of analysis and Figure~\ref{fig:thematic-map} presents the thematic mapping carried out to identify the main final themes.

\newpage

\onecolumn

\begin{table}[p]
\centering
\caption{Initial Coding Themes}
\label{tbl:themes}

\begin{tabular}{p{0.2\textwidth}p{0.75\textwidth}}
\toprule
\textbf{Initial coding themes} & \textbf{Example quote(s)/description} \\
& (n.b. some quotes may be coded under multiple themes) \\
\toprule
Fear of misuse of personal data by services & 
e.g. fear of mis-use of personal data by services staff: \newline 
‘It’s the privacy of my information. Who can get hold of that information, is it safe?’ (2004)  \\
\midrule
Theft of personal data & 
‘Because you hear a lot of stories, people just stealing people’s identity. That’s why a lot of the time I’m trying to think, “shall I do it? shall I not do it?”...that’s the one thing that I don’t like!’ (1050) \\ 
\midrule
Services ask for too much information & 
‘I don’t see why anyone else needs your date of birth, unless they’ve got their own reason to do something like a scam.’ (1049) \\ 
\midrule
Questioning why personal data is needed & 
‘If I’m applying for a credit card or something like that... why do you need to know the colour of my skin?’ (2006). \\ 
\midrule
Fear of machine bias & 
‘They say it’s anonymous. It’s almost that thing that if they know where you’re from, it could easily sway their decision, despite all the reassurances. You get reassurances that it’s not going to be given to anyone, but also, it’s unnecessary.’  (1045) \\ 
\midrule
Fear of human bias & 
e.g. Fear of discrimination: \newline 
‘As soon as you put, “Pakistani,” from there it’s like, “Right, okay, mate. Your chances are being brought down.” I’ve noticed that in Scotland.’ (3002) \\ 
\midrule        
Concern around longevity of data and accuracy & 
‘I think there should be some time limit where they should store it and then get rid of it or something. Maybe at that time, you discuss something personal, but then that data is everywhere, and bearing in mind, other people can access it straightaway.’ (4017) \\ 
\midrule
Lack of knowledge on when to opt-out & 
‘There is no information about what to do to keep safe online.’ (1040) \\ 
\midrule
Inaccessibly communicated terms and conditions & 
‘ I don't actually know why that question is asked. I don't get why it should make a difference. Why is that question asked? I don't know.’  (2006) \\ 
\midrule
Lack of knowledge on how to stay secure online & 
‘Those notices are hard to understand, full of technology jargon. I’m not fully able to understand the notices. I am anxious about the personal data and how it might be used. (3023, Interview notes) \\ 
\midrule 
Services lack clarity in communicating data sharing principles & 
‘I was a bit hesitant because I don’t know to what extent that information will be protected. It really held me back in not wanting to use my full surname, right. It held me back, and in some cases I wasn’t using my full surname... I’d rather not, because sometimes you don’t know where these things lead to, if you know what I mean.’ (3016) \\ 
\midrule
Experiences of scams and fraud & 
‘when I was online shopping. It was for a supplement, there was an advert for fat loss supplements and like I said it was during covid and was feeling really low and not happy with my weight. They took 200 pounds but I never got any money back or received any products. I paid with debit card and I was with Natwest and they did nothing about it even though I raised it with them. This happened last year. I was feeling low anyway and that really knocked my confidence and probably why I don’t do online shopping because I had that negative experience.’ (4016) \\ 
\midrule
Data labels do not match with my self-ID & 
‘I feel like, at any time I am given the option to, “prefer not to say”, I kind of tend to just go for that... I identify myself as probably British, but then I know that question is not really associated in terms of how I associate, it’s more in terms of my ethnic background and stuff like that.’ (2006) \\ 
\midrule
Information refused or refusing to use services  & 
e.g. Refuse to use/install smart meters: \newline 
‘I have this weird suspicion about smart meters that then they use it to rinse you more. I'm not sure if that's true. Often when they give you smart meters then they sign you up for automatic readings. For some reason I have this suspicion that that means you end up paying more. I don't know if that's actually true.’ (3017) \\ 
\bottomrule
\end{tabular}
\end{table}

\begin{table}[p]
\centering

\begin{tabular}{p{0.2\textwidth}p{0.75\textwidth}}
\toprule
\textbf{Initial coding themes} & \textbf{Example quote(s)/description} \\
& (n.b. some quotes may be coded under multiple themes) \\
\toprule
Withdrawal from online services & 
‘I feel nervous about my information. It make me feel anxious thinking about privacy and what people might be doing with my information or what kind of things people will find out and share with others. I think that’s what limits me from using online stuff more.’ (4016) \\ 
\midrule 
Fear of doing the ‘wrong’ thing online & 
‘I don’t like those cookie settings all the time. It makes me feel anxious about whether I might have pressed the wrong thing and how they will use my information if I press the wrong thing.’ (3023) \\ 
\midrule
Lack of trust & 
I’m aware of applications for benefits being sabotaged, small elements in the DWP [Department for Work and Pensions] department that were racist to prevent and de-lay applications amongst minority ethnic people. So, I’m very aware of how information could be used against you.’ (2014, interview notes) \newline 
‘I think we’re living in very untrustworthy, very cynical times, and all I’m trying to do is just protect myself and my son especially, because he’ll be more vulnerable with his disabilities, so that’s all I’m trying to do, really.’ (2023) \\ 
\midrule
Fear of being perceived as evasive  & 
‘Then at the same time you feel, if I don’t fill that form, will they think I’m trying to hide something?’ (2004) \\ 
\midrule
Selectively sharing or inconsistently providing information & 
‘I avoid sharing ethnicity information when I can, I’m not 100\% confident about where my information will be used and shared and if it will be taken in a positive perspective. I just say that I’m Black African Caribbean and Black other so it’s a bit purposefully vague.’ (2014, interview notes) \\ 
\midrule
Experiences of Racism (online) & 
‘I think it’s just a toxic environment and I’ve had experience of very negative, racist comments and abuse on social media.’ (3020) \newline 
‘I don’t think you’ll get treated any better. If somebody is a racist, they’re going to be a racist regardless of your data. It’s all a tick-box exercise. I don’t like those things without substance, things that don’t have a purpose.’ (1045) \\ 
\midrule
Experiences of Racism (offline) & 
‘Because I’ve been asked my name on many an occasion, I’ve been searched... I’ve been asked to do all sorts of stuff on many an occasion that is a breach of my privacy. A personal breach of my privacy. So, I’m reluctant to just give my name out willy-nilly or my details out willy-nilly. Even signing that document [participant consent form] that you’ve given me I was thinking twice about at one point, but I thought no, it’s for the greater good so it’s fine.’ (3020) \newline 
‘One of my daughters wears a hijab and the other ones don’t, and I still feel very like, oh, my gosh, I’ve got to make her grow a backbone because I don’t know how that’s going to affect her.’ (1045)’ \\ 
\midrule
Frustrations at profit motives linked to data collection &
‘Personal information about me has been disclosed without my consent due to collection of cookies and selling data. Not sure where it goes – we are giving our data for free and they sell it off and this is unsettling’ (3007, Interview notes) \newline 
‘I think they changed the right to your health data. So the information that the doctor has about you was now open to everybody for the government to sell off to private companies.’ (2010) \\ 
\midrule
Feeling powerless and forced to comply & 
‘Migrants that come in that are going through the asylum process or the refugee process, they’re too scared to say anything. They just want to keep their heads down. Not because they’re illegal or anything, but even if they’re illegal they still have rights too.’ (2007) \\ 
\midrule
Lack of choice – forced to comply & 
‘I think the value, if it’s used in a positive way, outweighs the negative because if you can point out there’s an issue here within this group of people that they should be aware of, and just let them know. Say, “Look, this is what we’ve found. We think the issue has stemmed from this”... but it should all be about choice really.’ (2010) \\ 
\bottomrule
\end{tabular}
\end{table}

\begin{table}[p]
\centering

\begin{tabular}{p{0.2\textwidth}p{0.75\textwidth}}
\toprule
\textbf{Initial coding themes} & \textbf{Example quote(s)/description} \\
& (n.b. some quotes may be coded under multiple themes) \\
\toprule
Lack of transparency & 
I could understand if for example you find that in diabetes there are more people from the African, South Asian descent who are likely to suffer diabetes, I would understand that. But there are some times whereby they are asking you of your ethnicity, and you’re thinking, why do you need that information anyway?’ (2004) \\ 
\midrule
Fear of surveillance and tracking across platforms & 
‘Also it's like when you're going shopping, you have those loyalty cards, before, because I worked in a supermarket at one time, before I knew, I used to - [...] - so you are using that, you're thinking, I will accumulate so many points, not knowing that they see your shopping habits. This is how you shop, and it's like big brother looking at you, what you are doing.’ (2004) \\ 
\midrule
Engaging in self-censorship & 
‘I think websites that dig into personal information more than is needed. ... Personal data most to me is like address, bank detail, disability – things that are not necessary for people to know, you know when it’s not the point of our discussion. I only tell you if I feel they need to know.’ (2017) \newline 
‘I try not to interact with people who have strong opinions online’ (1045) \newline 
‘I'm very careful because social media has shown all of us. We've seen to what extent social media can be very, very good, and can be bad. I am very cautious now. I'm very cautious. I try to limit what I say’ (3016) \\ 
\midrule
Lack of ‘human’ contact & 
‘They always tell me, 'Why do you walk in here? There's an app, why don't you do that?' I say, for me, when I am there, I can't express my urgency, I can't say my reason really, because most of the things are drop down, choose, so you can't - they give you the option of three or four and none of the options is what you want to say’ (2013) \newline 
‘I find it easier to speak to the GP face to face – feel more confident trying to speak than go online. So I phone the GP, first thing in morning, on hold for 30 minute sometimes, struggled to get an appointment sometimes. Nobody’s ever said that I should use the online service.’ (4011) \\ 
\bottomrule
\end{tabular}
\end{table}

\newpage 

\end{document}